# Machine learning techniques applied for detection of nanoparticles on surfaces using Coherent Fourier Scatterometry

D. K<span>OLENOV</span> *, S.F. P<span>EREIRA</span>

*Optics Research Group, Imaging Physics Department, Faculty of Applied Sciences, Delft University of Technology, Lorentzweg 1, 2628 CJ Delft, The Netherlands*
*\*d.kolenov@tudelft.nl*

**Abstract:** We present an efficient machine learning framework for detection and classification of nanoparticles on surfaces that are detected in the far-field with Coherent Fourier Scatterometry (CFS). We study silicon wafers contaminated with spherical polystyrene (PSL) nanoparticles (with diameters down to $\lambda/8$). Starting from the raw data, the proposed framework does the pre-processing and particle search. Further, the unsupervised clustering algorithms, such as K-means and DBSCAN, are customized to be used to define the groups of signals that are attributed to a single scatterer. Finally, the particle count versus particle size histogram is generated.

The challenging cases of the high density of scatterers, noise and drift in the dataset are treated. We take advantage of the prior information on the size of the scatterers to minimize the false-detections and as a consequence, provide higher discrimination ability and more accurate particle counting. Numerical and real experiments are conducted to demonstrate the performance of the proposed search and cluster-assessment techniques. Our results illustrate that the proposed algorithm can detect surface contaminants correctly and effectively.

## 1. Introduction

Much research on detection and localization of deep-subwavelength objects based on optical scattering has been done, covering a wide range of particle types such as viruses, bacteria, dust and nanofabricated features [1–4]. Regardless of the various approaches, the physical principle that underlies these studies remains the same. By analysing the light that is scattered to the far field after being reflected from a surface containing nanoparticles or other types of contamination, one tries to get information on the density, size, material of these nanoparticles [5]. In the context of the semiconductor industry, we can think of unwanted contamination on the silicon wafers in the nanometer-size scale. This contamination can occur at different stages of the lithography process, and it is important to check blank or patterned wafer as well as the mask (reticle) itself. The reticle quality and reticle defects continue to be a top industry risk [6]. To ensure the quality and high yield in semiconductor manufacturing, contamination due to isolated particles in the size range of from 20 nm to 1 $\mu m$ in diameter should be detected and, if possible, localised and removed.

The main techniques to study these nanometer-size features are scanning electron (SEM), dark and bright field microscopes. For electrically conductive materials, the surface analysis in the reflection mode is straightforward with SEM. If the scattering objects are buried inside the structures, transmission electron microscope (TEM) or scanning TEM (STEM) using a beam or a focused spot of electrons can be used [7]. With these techniques, sub-nanometer resolution can be achieved; however, it is hard to implement them in the production line, and generally, these techniques are considered to be slow. In addition, if relatively high beam current and acceleration voltage for the electrons are used, the analysis with SEM can also produce cracks on the surface or permanent thermal damage.

Subsequently, dark-field techniques, where only the scattered portion of the light is captured, are powerful tools for high-throughput analysis. The state-of-art systems work with bare wafers, smooth and rough films, and deliver defect detection sensitivity aimed at the 7 nm logic and advanced memory device nodes [8]. Since the direct reflected light is eliminated from the measured field, the incident power has to be high to produce enough scattering and sufficient signal to noise ratio (SNR). Hence, similar to SEM, in dark-field measurements, there is a potential to alter or damage the sample under study due to thermal effects [9].

Bright field techniques, where the reflected and the scattered light from the surface are measured, solve the issue of the sample damaging since they use very low incident power. However, similarly to dark-field, the small inherent scattering and consequently low SNR renders the limit of the sensitivity. In this context, it is hard to detect tiny particle sizes, with diameters < 100 nm in bright field mode using the visible wavelengths.

To solve this issue and to allow for the detection of such particle sizes, researchers have proposed various methods including interferometric ones such as label-free interference reflectance imaging of IRIS or ISCAT [10] and non-conventional sensing with optical forces in optical pseudo-electrodynamics microscopy (OPEM) [2]. Another family of techniques that are suitable to study nanotechnology materials in far-field is based on Quantitative Phase Imaging (QPI) [11]. The method of optical interferometric microscopy, in particular, has demonstrated an outstanding result in detecting 20 nm wide defects in patterned wafers [12]. A volumetric (3D) analysis for processing focus-resolved images of defects is enabled via the combination of scattered field optical microscopy and through-focus scanning optical microscopy. The results include the detection of sub-20 nm patterned defects [13]. Alternatively, one can obtain high sensitivity and low power of the illumination by measuring the light that is scattered from the particles to the far field in a smart way such that the SNR can be improved as compared to dark field techniques. This is the core of the technique used in this paper, namely Coherent Fourier Scatterometry (CFS): it is a low cost, robust, and suitable for the detection of polystyrene latex (PSL) nanoparticles down to 50 nm in diameter, and possibly even smaller ones [14–17].

For the detection of very small particles using CFS, it is crucial to optimize the entire system. Reliable numerical tools have been developed to understand the parameters that could influence the scattering process such as polarization, beam shaping, and how the data should be collected. Experimentally, besides a robust design, improvements directed to the detection system (such as noise suppression by introducing heterodyne detection system and beam shaping [14, 18]) have been implemented. At last, the data processing is of extreme importance, and this is the main subject of this paper.

The scatterometry data become useless if the algorithm that treats the data cannot effectively discriminate between different sizes of the particles present on a particular surface. One complicating factor is that, besides the inherent noise related to the detection of light, in a production environment, data can be corrupted with several other sources of noise and artefacts. In addition, the presence of extensive size-range contamination severely complicates the analysis of individual particles. In the worst-case scenario, if the measured data is examined in the wrong way, it can lead, for example, in the case of lithography, to a drop in system productivity. Finally, taking into account the growing amount of data, techniques such as CFS lack the tools of being able to process raw data sets automatically and effectively. Recently, to overcome the challenges of detection and classification of smaller particles, machine learning methods, including the regularized matrix-based imaging framework [19], principal component analysis [20], and convolutional neural networks [21] were applied to image-based defect detection.

The objective of this paper is to develop a full framework for particle size classification in scatterometry data consisting of pre-processing, signal search and histogram formation with an algorithm that can be directly targeted at data that is corrupted with noise and drift, as well as including mixed-size particles per sample. For this framework, we relied on the established noise-removal and unsupervised clustering techniques and adapted them to detect the nanoparticles. We developed a parametrized search by thresholding that picks the differential signal shape (raw data from the scatterometer) and relates it to the sought information (size distribution and location of the particles). By using these techniques, we show that the nanoparticles could be accurately quantified, even in the case of high densities. With sufficient resolution, a sample containing a mixture of nanoparticles with 60, 80 and 100 nm has been analysed in conditions where the data set had a lot of noise and drift due to the scanning issues related to the CFS tecnhique. Our framework enables the demanded automatic analysis of the scatterometry data and facilitates the validation of the detection results.

The paper is organized as follows. In Section 2, a brief overview of the measurement process and data is presented. Section 3 contains a description of sub-problems for the data set analysis. Section 4 describes the proposed algorithms of pre-processing, search, feature extraction, supervised clustering, and computational complexity. Section 5 shows the experimental results with the framework implementation and compares the accuracy of several classification algorithms incorporated in the scheme. In Section 6 and 7 we finalize the paper with discussions, conclusions, respectively. The summary of the functions used in this paper is given in Table 1.

Table 1. Glossary of the main functions used in the manuscript.

| name | explanation | mathematical description |
|---|---|---|
| detrend | Trend can be modeled and removed from the time series | minimize $J = \sum_x \left[y_x - (ax+b)\right]^2$ where measured data values $y_x$ in time $x$ and $a, b$ are parameters to be minimized |
| abs | Modulus of the real number | $\|y\| = \begin{cases} y, & \text{if } y \geq 0 \\ -y, & \text{if } y < 0 \end{cases}$ |
| ceil | Round up or round towards plus infinity | $y = ceil(x) = \lceil x \rceil = -\lfloor -x \rfloor$ |
| dist | The Euclidian distance between a point $x_n$ and $\mu$ | $\text{dist}(x_n, \mu) = \sqrt{(x_n - \mu)^2}$ |
| ind1 | Sampling points of the singal at which | $\text{argmax}_x y(x)$ |
| ind2 | the amplitude is maximized/minimized | $\text{argmin}_x y(x)$ |
| size | Size or cardinality of a set is a measure for the number of elements of the set $n$ | $card(y_x) = n$ |

## 2. Methods

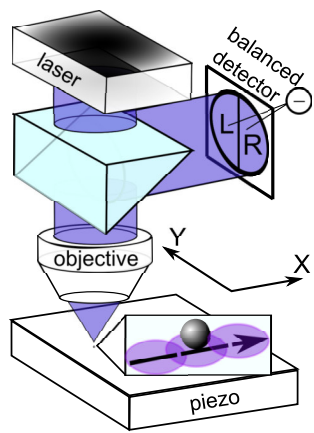

Fig. 1. Schematic of the experimental setup, showing the light path and the differential detection principle. To obtain the scatterometer maps, the sample is scanned in the x and y directions. For every X-Y position, the differential signal at the balanced detector is recorded.

Measurements with CFS are done via raster-scanning a ≈ 1 $\mu m$ tightly focused spot ($\lambda$ = 405 nm, $NA$ = 0.9) over the surface of interest. The Fourier plane of the objective obtained in reflection is imaged on the balanced detector. The sample is mounted on a 3D piezo-electric stage whose position can be controlled with sub-nm precision (P-629.2CD by Physik Instrumente). To reduce the amount of recorded data, and to increase speed and SNR, the Fourier plane is divided into two halves (perpendicular to the scan direction) and subtracted from each other using a balanced split detector (see Figure 1). In this way, for every X-Y scanning position, only one current value is obtained and stored as one point in the 2D scatterometry map. The differential detection allows having high SNR because the contribution from the rough background is minimized. If any clean part of the surface is analyzed, the acquired signal is virtually zero. For the areas containing particles, the spurious reflected light from the surface and light that is scattered due to the particle interferes at the detector. The total far field in the presence of the particle will become asymmetric as the particle is scanned through the focused beam. This implies that the left half of the field in the pupil is different from the right half, generating a nonzero photocurrent at the split detector, The recorded signals from the photodetector are the basis for the scattered maps (2D X-Y distributions) [22]. One of the significant advantages of the CFS approach is its high-sensitivity in localizing the centre of the particle in both transverse $XY$ and longitudinal $XZ, YZ$ planes. When the probe is focused on the interface, by scanning a spherical nanoparticle on the surface will render the so-called balanced pulse (positive and negative lobes of equal intensities, see Figure 2 B)). The zero-crossing of this pulse refers to the perfect alignment between the centre of the nanoparticle and the focused spot [23] (green point in Figure 2 B)). The effect of the defocus produces unbalance of the signal as well as a drop in the SNR drop, as it has been demonstrated in Ref. [23].

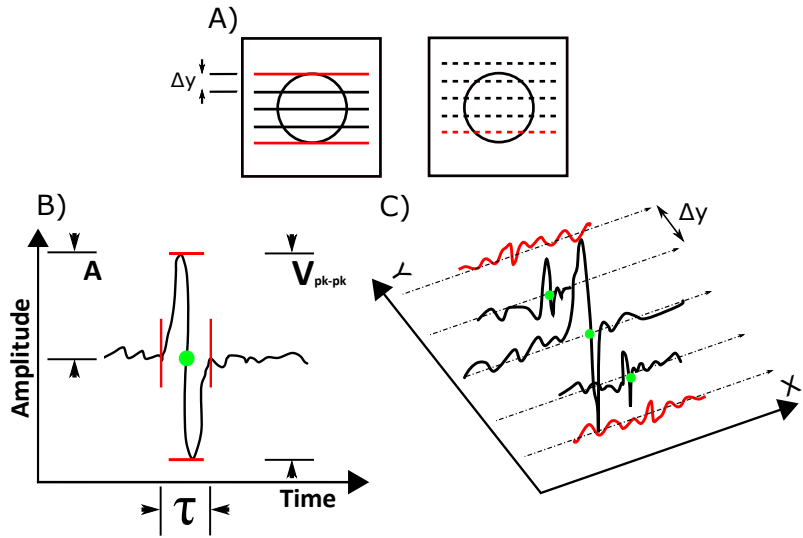

Fig. 2. A) 2D raster scanning procedure showing scan lines in the X direction separated by $\Delta y = 10$ nm in Y direction. The geometrical size of the studied particle is $d = 50$ nm. On the left - the first scanning line coincides with the edge of the nanoparticle, and consequently the differential signal will appear in 5 consecutive lines, with the red lines providing small amplitude of the signal. On the right - if there is an offset between the first scanning line and the edge of the particle, the signal due to this particle will be spread in fewer lines. B) A typical amplitude distribution of the recorded differential signal in time as one line containing the particle is scanned in the X direction - first maximum and then minimum when the positive lobe is equal to the negative lobe (balanced signal). The time axes is related to the X axis (length) by $t = v.L$, with $v$ being the scanning speed and $L$ the length of one scan line. Red lines constrain the features of width $\tau$ and amplitude $V_{pk-pk}$ of the differential signal. C) An example of a particle response as a collection of subsequent scans in X (separated by $\Delta y$) is called a scattered map.

When analysing the surface in a raster scanning fashion, one needs to choose the proper $\Delta y$ displacement step between the parallel lines of scanning. The bigger the step, the lesser the time it takes to cover the complete area, but the downside is that particles can be missed. The simplified picture showing the relationship between the scanning step $\Delta y$ and the particle size is shown in Figure 2 A. Naturally, if one wants to detect contamination of e.g. 50 nm, the step between lines of the scanning should be set lower than the particle diameter, for instance $\Delta y = 10$ nm. For relatively big particles, one can expect that every time the probe interacts with the particle, a high-enough scattering will be produced and thus the estimate for the amount of lines where the particle is visible equals to $n = d/\Delta y$, with $d$ being the diameter of the particle. N.B. the SNR is the defining factor for the effective amount of lines and the influence area of the particle is bigger than the physical size as according to its scattering cross-section. Yet, the outlined picture highlights the idea that there is a certain minimum and maximum expected number of beneficial signals that would emerge from using different settings of $\Delta y$. For tiny particles < 80 nm, using the wavelength $\lambda = 405$ nm, one can expect that the scanning lines that would go through the edges of the particle have much less SNR (see Figure 2 A and C), because the amplitude of the scattering becomes small as the focused spot goes away from the center of the particle. Furthermore, mismatches between the position of the particle and the scanning step might occur (dashed lines Figure 2 A) and in this case, even fewer scan lines containing signals due to the particle are obtained. The rule of thumb is to have at least two signals that come from an isolated particle that is distinguished from the background. Finally, the nanoparticles are generally classified based on their dimensionality, where the size of the calibrated sphere is associated with the features of $V_{pk-pk}$ amplitude and time-width $\tau$ of the measured differential signal (Figure 2 B). The time axes is related to the X axis (length) by $t = v.L$, with $v$ being the scanning speed and $L$ the length of one scan line.

## 3. Sub-problems

The task of detection and classifying the particles using scatterometry data can be split into sub-problems. In this section, we discuss these sub-problems: pre-processing of data, finding the particle-like signals, estimation of the width, cluster assessment.

### 3.1. Pre-processing

The goal of the pre-processing task is to prepare the raw sampled data for further steps. Commonly, a DC bias and sometimes baseline fluctuations in the signal at the detector can occur due to vibrations and other experimental factors. For the removal of the various electronic noise, low-pass (LP), notch filtering and wavelet-based subtraction were applied.

### 3.2. Selection of suitable Amplitude and Width

We use two parameters for the object detection: the $A$ amplitude ($V_{pk-pk}/2$) and the $\tau$ width of the complete differential signal due to a single particle in the time-domain (see Figure 2 B). We look for an algorithm that is robust to non-particle signals that can be present in the data. Examples of such signals include environmental vibrations or large defects present on the surface of the material, which we can consider as false-detections.

In the scan direction $X$, multiple particles may be present on one scan line since the density of particles can be high in some areas of the surface. Multiple maxima and minima needs to be determined on one scan line, sorted and the relative distance between different signals needs to be determined. Each "transition" between maxima and minima is associated with the corresponding zero-crossing position at the middle of the signal. Finally, fine adjustment is needed to define the width of the particle-like signal accurately; this is done by parametrizing it such that it can be distinguished from noise or another signal in the data set. Next, one needs to take care of the particle signal appearing at the border of the scan line. In this case, if the signal was sampled for one of the two lobes (positive or negative), the algorithm should estimate the complete width of the pulse.

### 3.3. Multiple line particle detection identification

A particle-like signal is distinguished from a false detection if the centroid of the signals (position of the zero between maxima and minima) have the same X position over multiple lines (see Figure 2 C). A false detection is identified when the particle-like signal is observed in only one scan line, and is further removed from the data. Finally, per signal group, the most clear particle-like signal and its features (see Figure 2 C) are stored for the histogram. The pulse with the biggest $V_{pk-pk}$ is a good representative because it corresponds to the centre of the particle in the X and Y directions.

There are many well-known algorithms for cluster determination, such as hierarchical clustering, K-means, DBSCAN [24–26]. Almost every clustering algorithm can be tuned to penalize one error more than the other according to the requirements. For instance, we can use the predefined vertical step of $\Delta y$ and set the expected number of zero-crossings associated with a single particle. Additionally, the clusters of zero-crossings (pair of X and Y coordinates) can have a characteristic spread of $\sigma$.

## 4. Algorithm

In this section, we show the specific tools and algorithms we have used to solve the sub-problems showed above. We also mention the computational efficiency and some other aspects of the algorithms.

### 4.1. Pre-processing

Among various noise sources that might be present in our experiment [14], the power line interference and the baseline wandering can strongly affect the further detection and classification of particle signals. The 50-60 Hz local power-line frequency (bandwidth of < 1 Hz) can be mostly removed by analogue hardware during acquisition, and the remaining noise is removed digitally using the notch filter. However, the baseline wandering is not easy to be suppressed by analogue circuits. Hence, we take the notch-filtered waveform and subtract the wavelet decomposed version of the same signal to recover the clean particle signal (more details in Appendix: The pre-process filters). This step effectively introduces the point by point correction to the wandering profile. Finally, an average filter (LP) is applied to remove glitches. This approach can be considered as more rigorous because it relies on the sampling frequency used in the experiment. The routine is based on a contribution from reference [27]. A less accurate way of dealing with the offset in the data can be MatLab's *detrend* function that removes the best straight-fit line from the data in a vector of the sampled points.

### 4.2. Selection of suitable Amplitude and Width

Hyperparameters:
$$A, \tau, NullingR, Window_y$$

These are user-defined parameters of expected threshold amplitude $A$ and width $\tau$. Since the multiple expected amplitudes and widths are passed iteratively, the results from the previous search should not translate to the consecutive one. Let's consider the 2D measured data $\boldsymbol{I}_{ij}$ with each row representing a single scan line (Y) and column representing the sampling point over the width (X of the Figure 2 C). The differential signal at the detector for the $i = 4$ scanning lines and with $j = 4$ samples in horizontal direction of scan is given in Eq. 1

$$\boldsymbol{I}_{4x4} = \begin{bmatrix} I_{11} & I_{12} & I_{13} & I_{14} \\ I_{21} & I_{22} & I_{23} & I_{24} \\ I_{31} & I_{32} & I_{33} & I_{34} \\ I_{41} & I_{42} & I_{43} & I_{44} \end{bmatrix} \tag{1}$$

The parameters of $NullingR$, $Window_y$ represent the half-width and length of the region to be zeroed w.r.t reference sampling point. Hence, for measured data, if $I_{33}$ is the reference position (centre of the particle), the $NullingR = 1$ and $Window_y = 3$ dataset becomes:

$$\boldsymbol{I}'_{4x4} = \begin{bmatrix} I_{11} & I_{12} & I_{13} & I_{14} \\ I_{21} & 0 & 0 & 0 \\ I_{31} & 0 & 0 & 0 \\ I_{41} & 0 & 0 & 0 \end{bmatrix} \tag{2}$$

Thus by, $NullingR$ and $Window_y$, the user can zero the lines that are close to the reference detected particle. Per line, the algorithm looks for the multiple peaks and minima and checks whether their absolute values fall under the amplitude $A$. The retrieval of secondary peaks and minima allows increasing the overall accuracy of the algorithm. By default, we assume every particle-like looking signal to be a forward signal (Figure 2 B)). The reverse signal can be stored separately or included in the estimation process. Some key reasonings are highlighted in the following bullet-points and also shown in Figure 3.

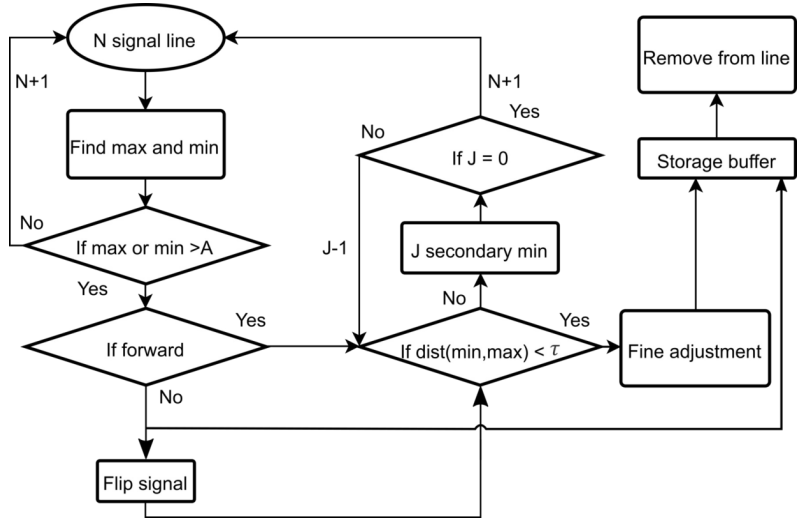

Fig. 3. Block diagram of the signal search algorithm that starts with the $N$ signal line.

- Find and store the values and indices of global line maxima $ind1$ and global line minima $ind2$. Next, check the amplitude condition $abs(max1) > A\ OR\ abs(min1) > A$. Store True of False for the first condition.

- Define whether the signal is forward (maxima appears before minima), choose between ignoring or flipping the reverse pulses. Check whether the distance between 2 indices $abs(ind2 - ind1) < \tau$ fits the condition of the time-width. Store True of False for the second condition.

- When both conditions are true, a particle is roughly detected. We calculate the position of the particle's signal zero-crossing by taking the average between maxima and minima position $middle = ceil(abs(ind1 + ind2)/2)$ (ceil function rounds towards plus infinity), perform the fine adjustment (next section) and remove the signal from the data set. The $NullingR$ is global parameter that represents half-distance in indices to replace with zeroes about the $middle$ of the signal. The rule of thumb, in this case, is that region to be zeroed should not exceed the time width of the particles you are looking for.

- If only the amplitude condition is satisfied, the indices of multiple minima (above threshold) that belong to the current line are checked to fall closer to the $ind1$ than $ind2$. If other minimum falls closer, reassign the $ind2$ and repeat the width check. If both conditions are satisfied, remove the signal from the data set and apply the $NullingR$.

- The multiple particle search routine is to find numerous maxima (above threshold), *sort* them in descending order (see Alg. 1), and, maximum by maximum, follow the steps outlined previously. If there are multiple particles on a single line, the algorithm returns X's corresponding to the particles middles.

Throughout this paper, we will use the terms "zero-crossing" and "middle" interchangeably, following the variable name of *middles* as defined in the MatLab software.

### 4.3. Fine adjustment for the boundaries of particle's signal

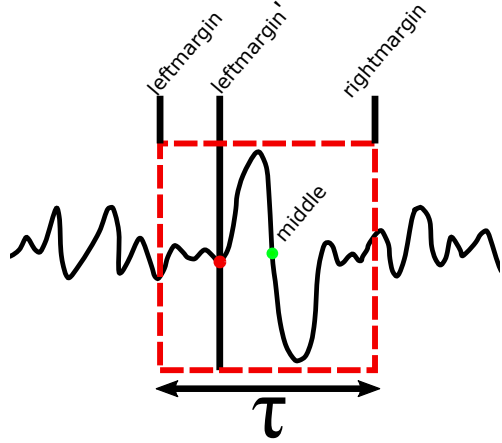

Fig. 4. A sketch showing the margins of the signal separating it from the background. The fine adjustment algorithm is to go from *leftmargin* to *leftmargin'*.

Fine adjustment is the part of the search process right after the width, and amplitude conditions are satisfied. We assume that from the *middle* position, the particles' signal occupies the same amount of samples on both sides of the signal (spherical object). The initial guess for the left margin of the signal is $leftmargin = middles_{ind} - \tau/2$. To make sure that the signal doesn't go outside the indexing in Matlab, i.e., $middles_{ind} - \tau/2 <= 0$, we rewrite the left margin as index 1. In this case, in the procedure that follows, we should rely on the *rightmargin* to be defined accurately and then *leftmargin* is recomputed based on it. Analogously, the right margin is calculated as $rightmargin = middles_{ind} + \tau/2$. If a signal appears close to the right border, we rewrite right margin as the last index of the sampled voltage vector. The crucial part of the fine-adjustment step is to cut-out the region of signal for zoom-in study, i.e. from *leftmargin* to *rightmargin*. The secondary minima of the cut-out region are checked to fall closer to the middle position, compared to the initial *leftmargin*. If there is a closer point, it is redefined as *leftmargin'*. The reason for this is an observation that typically there is a small dip in the signal preceding the quick rise in amplitude of the particle pulse. Further, we reassign the *rightmargin* of signal to be the same separation as to the left $rightmargin = middles_{ind} + abs(middles_{ind} - leftmargin')$. Finally, the $signalSize = abs(rightmargin - leftmargin')$, and one can notice that in our procedure the estimator can generalize outside the original size of the sampled vector.

### 4.4. Clustering of data from one single particle

The steps of the algorithm presented previously result in an array of coordinate pairs for **(X,Y)**, that correspond to the position of the *middles* (zero-crossings) of each signal and has dimensionality $2 \times N$, where $N$ is the number of *middles*. Since one particle results in a few signals at consecutive lines of the scan, one should recognise a group of the particle-looking signals as a centroid that represents this specific particle. In this way, the particle-size distribution histogram will identify one particle on the sample corresponding to one cluster of signals. The centroid of the cluster will correspond to the line with the highest $V_{pk-pk}$ of the cluster, and consequently the particle's center.

We modify a well-known machine learning algorithms of K-means and DBSCAN to recognize the clusters of the particle-looking signals and use prior information that can help to spot the isolated particles. One complicating factor that can be present in the data is the random drift between the lines when sampling. The drift manifests itself in the shift of the signal zero-crossing (see Fig. 2C) position in the *X* direction between consecutive lines. In the Appendix Modified K-means, DBSCAN and comparison, we define several algorithms that can account for the drift in the data. We compare the computational complexity of the modified K-means to the algorithm of DBSCAN. Besides, we highlight the sensitivity of algorithms to initialization parameters.

## 5. Results

Throughout this section we experimentally study three different samples of the PSL particles spin-coated on the silicon wafer. The first sample includes particles with diameters of 50 nm, the second 100 nm, and the third a mixture of 60, 80 and 100 nm. The details on sample preparation are outlined in the Appendix: Preparation of the samples.

### 5.1. Pre-processing and search.

In the high-scale IC manufacturing, typically double side polished wafers are used. The block of pure crystalline silicon is diced and polished right before the deposition of the resist. Due to the lack of precision in the wafer holder, unstable rotation and heat deformation, the polishing can affect the flatness of the wafer. Additionally, the thickness of the wafer is not uniform across the sample [28]. This effect mostly occurs at the edges of the wafer. Nevertheless, the scanners need to provide information over the entire wafer under the study. For sensing or particle detection applications using CFS, the probing light should be focused on the interface between air and top surface. Due to several experimental factors during the scanning, the baseline (differential signal when no particle is present) may fluctuate or drift from the expected zero value. Hence, occasionally, the data set might include DC offsets mixed with low frequency noise (baseline wadering) [29, 30]. This problem can be corrected as shown in the data presented in Figure 5 A) raw data (top), and with baseline correction (bottom).

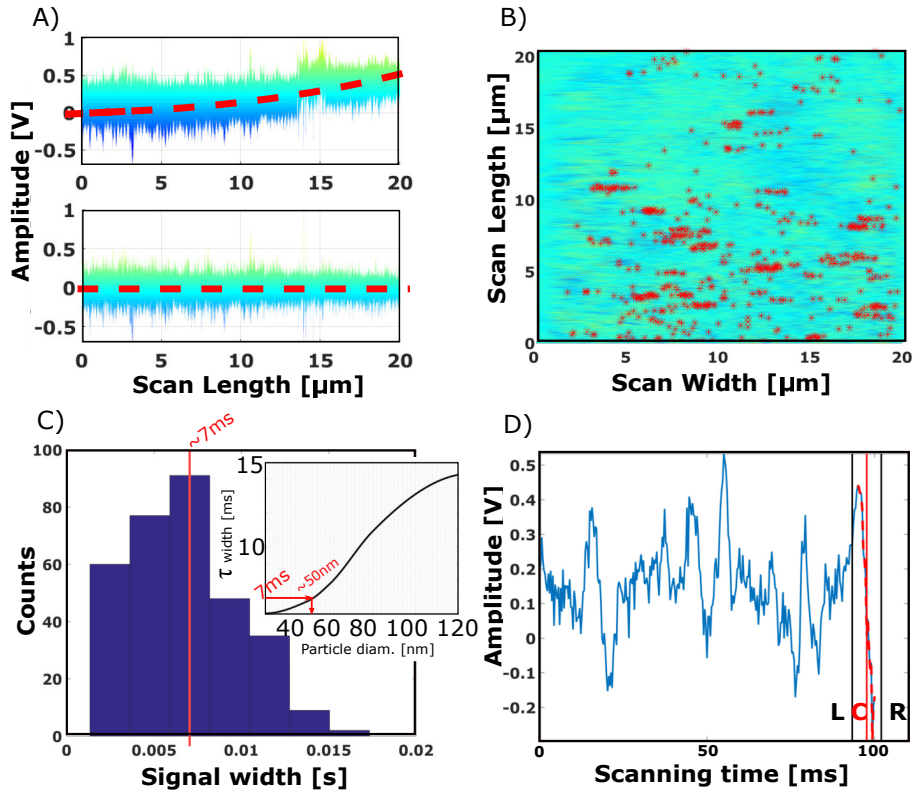

Fig. 5. A) Top - the side view (along Y) of the raw sampled data wherein the baseline wandering is present. Bottom - the corresponding data after the baseline wandering is removed. B) Top view on the same scan with the red points representing the detected zero-crossings. C) Histogram representing the particle size distribution, based on time width $\tau$ from detected pulses. The inset shows the calibration of size of the particle as a function of the time width of the signal. D) Example of the line from the data set, the dashed line is an initial guess for the time-width and left - L and right -R boundary is returned from the fine-adjustment step. The scan speed per line is such that a scan width of 20 microns in X takes 100 [ms].

Further, the scattered map from the bottom Figure 5 A) is analysed with the search algorithm (Section 4.2) to produce the corrected data that is seen on the 5 B). Here we analyzed a random area from the calibrated sample and the histogram nicely peaks at the position of the $\tau = 7.05$ [ms] that corresponds to PSL particle with 50 nm in diameter (see inset with the callibration curve) that agrees with the recipe of the first sample. For the area that contained only a few particles, one can notice a relatively high amount of counts, and this is because all the localized zero-crossings contribute to the output histogram. The SNR ratio for this dataset is low $SNR = 7.14$ [dB] while the algorithm can still localize the particle detections, including the one that resides at the border of the scan, thus generalizing beyond the input data.

N.B. The particle classification in CFS is based on the width of a time-domain particle signal. The quantitative limit of the post-processing framework for discrimination between the different-size particles is defined by the accuracy of the fine-adjustment routine of Section 5.1. More specifically, in the ability to find the minima closest to the rising edge of the differential signal. If we assume the infinite sampling of the signal and low noise, there are virtually no limitations on how accurate the position of minima can be defined, aside from those emerging from the numerics or computational effort [31]. On the practical side, there is a limitation in the manufacturing of the monodisperse PSLs. The target size of the particle diameter has the uncertainty in the range of 1 − 2 nm [32].

### 5.2. Comparing the accuracy of clustering routines on a data set with drift.

The source of the drift originates from the sampling at the detector being asynchronous process with respect to the piezo stage movement. When the piezo controller passes the initialization signal to the computer, the jitter and USB connection produce a random time delay before the sampling will actually start.

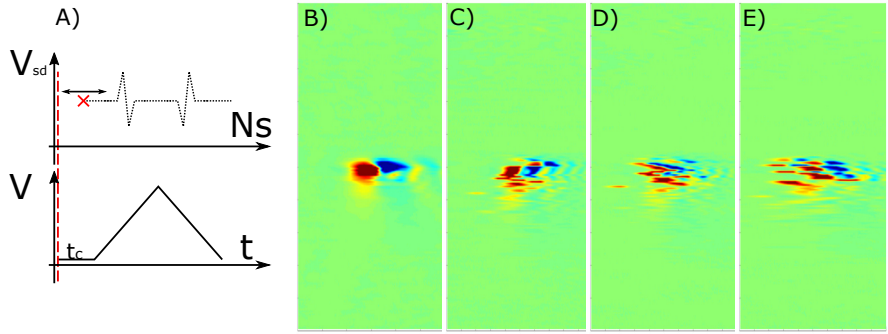

Fig. 6. A) The sampling done asynchronously, and the start sampling point (red cross) has fluctuation in time - above. The primitive of the voltage waveform for moving the piezo along one axis forth and back - below. The time-constant $tc$ tries to match the start of the piezo movement (uprising edge of the waveform) with the beginning of the sampling. Example of the isolated nanoparticle with different amount of drift present. The non-drift image B), an increasing amount of drift from 100, 90, 50 ms scanning time per line, C), D) and E) correspondingly.

One can mitigate the problem by introducing the constant waiting time $tc$ (empirical estimate) at the piezo before the voltage will be increased (Figure 6 A)). Yet, the random nature of the delay will not be equal to the introduced $tc$. When faster scanning is performed, the drift in the data set gets worse. Figure 6 B) - E) shows the same isolated nanoparticle scanned at a different speeds: 100, 90 and 50 ms per line, demonstrating an increasing amount of the distortion in the data set.

We take the data corrupted with the drift and compare the accuracy (Eq. 3) of two clustering algorithms as the average result of 100 random initializations.

$$Accuracy = 1 - \left|1 - \frac{N_{det}}{N_{true}}\right|', \qquad (3)$$

where $N_{det}$ is a number of detected clusters, hence isolated particles, and $N_{true}$ the actual amount of particles on the sample. This formula ignores the difference between the over- and under-estimate in the $N_{det}$. We will use the non-drift corrected "image" as a ground truth for this comparison providing us the number for $N_{true}$. The non-drift "image" is achieved by establishing a new synchronization approach with the trigger pulse generated at the piezo controller through analogue output upon each beginning and end of the scanning line.

The first test is to use some of the global parameters such as $n_{min}, \epsilon, n_{max}, \sigma_{thresh}$ according to the reasoning outlined in the Section 4.4 and Appendix: Modified K-means, DBSCAN and comparison. Recommended hyperparameters come from showing the program once how the "good cluster" looks like. A number of 100 random initializations were needed to get an idea on how the K-means algorithm will suffer from random initialization, specifically the starting number of clusters $K$ and their positions are randomly initialized. On the contrary, the DBSCAN, regardless of initialization, always converges to the same result (see Figure 7).

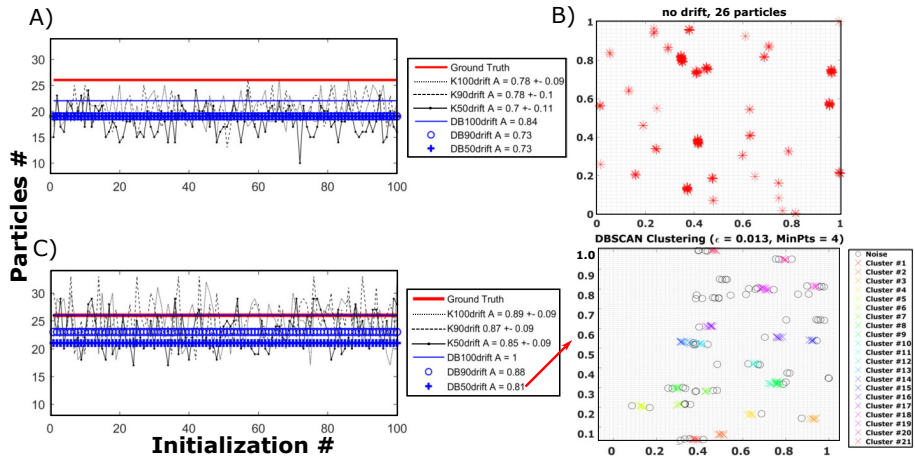

Fig. 7. The true number of particles (ground truth) in red. Comparison of the DBSCAN (in blue) and modified K-means algorithms (in black) for the three levels of drift present. 100 drift represents the least distorted data set, 90 drift dat aset with average distortion, 50 drift is the data set with severe distortion. Recommended in A) and tuned hyperparameters in C). Result of 81% accurate convergence by the DBSCAN for the case of 50 drift present B).

This test reveals that both algorithms can achieve relatively high accuracy > 70%. As it has been expected, accuracy on the data that contains less drift is higher and contains less uncertainty. On average, the accuracy does not exceed 84% for the case of the DBSCAN and the algorithm produces the same amount of clusters at every iteration. When the input data is shuffled, the only "non-deterministic" behaviour is in the label for the cluster being assigned, but not the composition of the cluster itself. The behaviour was firstly highlighted in the original paper of DBSCAN [26] where the authors claimed that convergence result is independent of the order in which the points of the database are visited expect the "rare" situations. This "rare" situations occur when border points belong simultaneously to two clusters. This border point will be assigned to the cluster that is considered first to avoid the overlap. In other words, there is always the same amount of density-reachable points from a reference point, hence the same amount of the assigned clusters is constant.

In the next test for both algorithms, the global parameters were manually adapted to yield higher accuracy (Figure 7 C)). The adjustments to the $K$, the desired number of clusters in K-means can be set higher than the elbow method recommends, and for the DBSCAN algorithm the $\epsilon$ parameter is crucial. This test demonstrates that with the aid of completely manual tuning, higher accuracy > 80% for any type of data set can be achieved. Even more, the $\epsilon$ parameter in the DBSCAN can be chosen to recover the 100% accuracy on the data set with the minor drift. N.B. The average convergence time for the DBSCAN algorithm is 0.01 second and for the K-means algorithm 47 seconds on laptop Dell Inspiron 7577.

### 5.3. Benefit of the centroids re-assignment

For the domain of the semiconductor industry, specifically for the lithography process, it is crucial that cleaning can be performed if contamination above a certain size is present on the sample. In this way, for instance, the very small particles are of minor importance for the pellicle layer above the UV mask, and only if the bigger particles are present, cleaning action needs to be taken. In the absence of the pellicle, on the contrary, one should take care only about the small contamination landing on the mask [33]. The quantitative description of the surface, provided by the surface scanner in this regard becomes very crucial. The confusion between the different sizes of the scatterers on the sample should be minimal. For our system, the width of the signal changes as scanning through the spherical particle is performed : it is highest when the scan line passes through the center of the particle and it is smaller in consecutive lines around the particle's center, as shown in Figure 8.

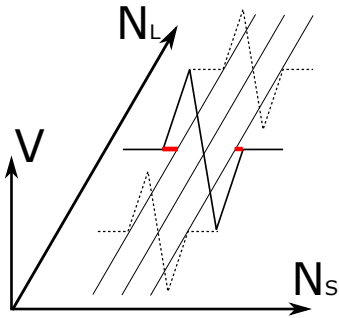

Fig. 8. Sketch of the signal from an isolated spherical particle visible over three consecutive line scans. The red region represents the increase in the $\tau$ width of the signal when the scan line passes through the centre of the particle as compared to other consecutive lines $\pm \Delta Y$ (signals as dotted lines).

In the first approximation, all detected signals can be fed to the histogram as it was done in Section 5.1. This approach would work properly if the data set would include a single particle size or if the contamination is reasonably different. Realistically, samples contain a wide range of particle sizes. If the pulses on the edges of the particle scan are included in the estimation histogram, they will contribute to the interclass confusion (classes represent diameters). In the Figure 9 we demonstrate the outputs from the signal search algorithm and corresponding clusters defined by the DBSCAN algorithm. This algorithm was chosen since the convergence time is faster than the modified K-means, and it had achieved higher accuracy at the previous test. The region of the sample under study is a good representative of the multi-class sample where additionally to the nominal 60 and 80 nm PSL particles, there are isolated particle-looking signals that are treated as outlier by the algorithm as well as the contamination of bigger particles $\approx$ 100 nm in diameter. The results of convergence by K-means algorithm for the same data set is presented in Appendix: The re-assignment of signal centroids by the K-means algorithm.

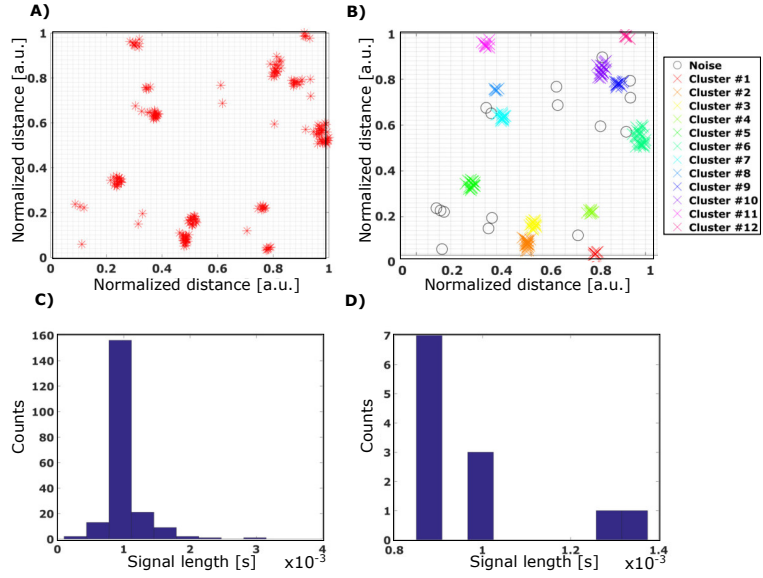

Fig. 9. A) The zero-crossings of the differential signals by the search algorithm and B) the corresponding isolated particles by converged DBSCAN. The data set includes minor drift where one scanning line of $\Delta x = 25 \mu m$ takes 100 ms. C) Histogram when all particle-looking signals are taken into account and D) when the signals corresponding to one particle are clustered and only one signal (highest $V_{pk-pk}$) is taken into account to represent one particle detection.

The first approximation histogram includes the side detection from the class of the 100 and 80 nm contributing to the class of 60 nm as well as features between the classes and it seems that there is only a single class present in the data (see Fig. 9 C). When all signals that corresponds to one particle are clustered and only the highest $V_{pk-pk}$ pulses from each cluster is assigned as being one particle (Fig. 9 B), we observe three separable classes in the histogram (Fig. 9 D), showing that this strategy solves the problem of particle size confusion.

## 6. Discussion

The approach of clustering the data has a downside, namely, the risk of losing the beneficial signals that correspond to very tiny particles. These particles may produce only one or two scan lines containing signals with sufficient SNR, if the selected step between the scan lines $\Delta y$ is too big. To improve the sensitivity of the algorithm even further, a separate routine could reconsider the outliers. This step could include adding a collection of matched filters operating in the time domain to filter out signals with the expected duration. Alternatively, one can try to establish spectral differences between the particle and non-particle signals (multiple wavelength approach).

While this study considered the detection of polystyrene particles, the technique could also be applied to extract features from a measurement of particles of different materials. Scatterometry is not an imaging technique, and some other features (such as material) can be recovered if one can model them and obtain more diversity in the experimental data. For example, instead of only looking at the time spam of the particle signal (related to the size of the particle), one can add its magnitude, which is proportional to the diameter and material of the particle [34]. Another example is the work of Potenza et al. [35] where using similar technique, they were able to recover the complex index of refraction of the particles, and in this way, reveal their material.

DBSCAN can yield higher accuracy than the K-means subroutine in the case when the scale of the data is well understood. Also, the convergence of DBSCAN algorithm is fast. Nevertheless, there is still room for implementing the K-means routine because the sensitivity to the hyperparameters is much higher in case of the DBSCAN, including the complete failure in defining the clusters from the initial data (see the Appendix: Modified K-means, DBSCAN and comparison). The K-means, on the contrary, can be considered as a more robust algorithm that yields relatively high accuracy, and at any initialization will always define a certain amount of clusters. The K-means algorithm is scalable to large data sets while the DBSCAN can suffer from the curse of dimensionality [36]. A final point to consider is when working on data sets with severe drift due to scanning, high density and wide range of the particle sizes, the DBSCAN can fail to cluster data [37].

Throughout the IC manufacturing process, large amounts of data need to be mined in a fully automated mode [38]. With a growing amount of data, we can envision that the line-by-line analysis of the data set can become computationally slow. Also, the total amount of hyperparameters is significant. Search and clustering routine in total has up to 6 parameters fed by the user. Hence, in future work, we will explore the potential of methods for handling big data, such as deep learning and CNN [39, 40].

## 7. Conclusions

We have demonstrated that Coherent Fourier scatterometry is capable of generating the 2D maps with the locations and sizes of PSL nanoparticles on a surface, down to particles with a diameter of 50 nm using low power illumiantion wavelength of $\lambda$ = 405 nm (on the substrate, the input power is $P$ =~ 0.026 mW). CFS relies on differential detection to minimize the contribution from the rough background, and uses photocurrent measurement in a raster-scanning regime generating a wealth of 2D data sets.

In this paper, we have developed a generalized framework that accurately extracts features of the differential signal produced by the scattering of a nanoparticle and uses these features for particle location and size determination. We have combined pre-processing with search algorithms based on the thresholding, such as peak-to-peak amplitude, and the width in time of the signal. The proposed method makes use of unsupervised clustering techniques to separate particles with high density on the samples. We adapt algorithms of DBSCAN and K-means and use them together with the simple prior.

We have tested the framework for data sets with high density of the particles, in the presence of large experimental noise and drift. The accuracy of the algorithm resulted in the 84% for the hyperparameters set semi-automatically, and the 100% accurate result for manually-tuned parameters. The algorithm of DBSCAN is a go-to solution because it works much faster than K-means. However, the latter is more robust because it is less sensitive to the change on the input parameters.

Finally, we would like to stress that while we tested the framework for the particular case of experimental data obtained with CFS, this method can be generalized to other experiments that involve measurements with differential detection, such as coherent time-addressed optical CDMA systems [41] and ferromagnetic resonance spectrometers (VNA-FMR) [42]. In these techniques, the data set might include mechanical vibrations or other experimental fluctuations, similar to the drift studied in this paper. We believe that the proposed framework is an essential addition to the nanoparticle detection experimental community.

## Appendix

*The pre-process filters*

The input-output description of the filter operation on an input signal vector $x(n)$, where $n$ is the number of samples, can be expressed in the form of the difference equation:

$$a(1)y(n) = b(1)x(n)+b(2)x(n-1)+...+b(n_b+1)x(n-n_b)-a(2)y(n-1)-...-a(n_a+1)y(n-n_a), \quad (4)$$

where $n_a$ is the feedback filter order, and $n_b$ is the feed-forward filter order. We design a second-order notch digital filter, thus $n_a = n_b = 2$ and Eq. 4 becomes:

$$a(1)y(n) = b(1)x(n) + b(2)x(n-1) + b(3)x(n-2) - a(2)y(n-1) - a(3)y(n-2), \quad (5)$$

with the notch at frequency 50 Hz and a bandwidth at the -3 dB level (q-factor of 35), we have angular frequency $W = 50/(f_s/2)$ and bandwidth $BW = W/35$. Thus, for sampling frequency $f_s = 3$ kHz, the coefficients are $\mathbf{b} = [0.998, -1.986, 0.998]$ and $\mathbf{a} = [1, -1.986, 0.997]$. The local power line frequency is removed from the data set and $y_{notch} = a(1)y(n)$.

Further, we subtract the wavelet decomposed version of the signal $y_{wd}$ from the filtered waveform $y_{notch}$ effectively introducing the point by point correction to the profile. The discrete wavelet transform (DWT) of signal $y_{wd}(n)$ is defined as a combination of a set of basis functions:

$$y_{wd}(n) = \sum_{k=-\infty}^{\infty} c_j(k)\phi_{j,k}(n) + \sum_{j=1}^{J}\sum_{k=-\infty}^{\infty} d_j(k)\psi_{j,k}(n) \quad (6)$$

Where

$$\phi_{j,k}(n) = 2^{j/2}\phi(2^j n - k)$$
$$\psi_{j,k}(n) = 2^{j/2}\psi(2^j n - k) \quad (7)$$

In Eq. 6, $\phi_{j,k}(n)$ is the scaling function, $\psi_{j,k}(n)$ is the wavelet function, $c_j(k)$ are the scaling and $d_j(k)$ are detailed coefficients. In this paper, the Daubechies 6 scaling and wavelet functions were used because it has been proved to be excellent in analysis of signals that contain baseline wandering [43, 44]. For computing the $c_j(k)$ and $d_j(k)$ coefficients, the low-pass (LP) and high-pass (HP) filters are being recursively applied to a signal. When the signal is processed for the first time, the HP filtered data gives the details and LP filtered data gives the scaling coefficients at level 1. The more times the filters are applied, the more detailed levels of the signal representation can be achieved. In this paper, we have used the decomposition level of $j = 10$ and have applied the translation factor of $k = 8$ for the scaling and wavelet function. The baseline wandering is removed and $y_{bcor} = y_{notch} - y_{wd}$.

Finally a simple moving average filter is applied according to Eq. 8.

$$y'_i = \frac{1}{M}\sum_{j=0}^{M-1} y_{bcor}[i+j]. \quad (8)$$

The output signal $y'_i$ is a result of averaging the points in the input signal $y_{bcor}$, and $M = 5$ is the number of points used in the moving average.

*Modified K-means, DBSCAN and comparison.*

Given, for instance, the *middles* coordinates **(X,Y)**, K-means clustering, can converge to a $K$ amount of clusters, among which per cluster we know the distance between each point and the position of the cluster centroid mean $\mu$. The first two algorithms are used to treat the outliers in the clusters $K$:

**Algorithm 1** Sort in descending order

```
1:  procedure SORT(X_n)                            ▷ Where x_1, x_2, ..., x_{jjj} ∈ ℝ
2:      inversions = 0
3:      for i in 1 : n do
4:          for jjj in 1 : i do
5:              if X(i) > X(jjj) then
6:                  inversions = X(i)
7:                  X(i) = X(jjj)
8:                  X(jjj) = inversions
9:              end if
10:         end for
11:     end for
12: return X
13: end procedure
```

**Algorithm 2** Remove if $\sigma$ improves

```
1:  Input: (X, Y)_{jjj} ∈ K           ▷ In cluster each middle (X, Y) is associated with dist to μ
2:  Output: Nremove
3:  set Nremove to zero
4:  Sort the points by dist                                  ▷ Descending order Alg. 1
5:  for b ← 2 to jjj_{dist} do
6:      m ← Eq.(9)
7:      if m < 0.1 then
8:          Nremove ← b − 1
9:          return Nremove
10:     end if
11: end for
```

Where the metric $m$, standard deviation $\sigma$, and average $\mu$ are computed according to Eqs. (9 - 11). For a random variable vector M made up of N scalar observations,

$$m = \frac{\sigma(X,Y)_{jjj}}{\sigma(X,Y)_{jjj-1}} - 1 \qquad (9)$$

$$\sigma = \sqrt{\frac{1}{N-1} \sum_{i=1}^{N} |M_i - \mu|^2} \qquad (10)$$

$$\mu = \frac{1}{N} \sum_{i=1}^{N} M_i. \qquad (11)$$

Example of such an algorithm (Alg. 2) applied to an arbitrary cluster is shown in Figure 10 A). The idea is to remove the points that are too far from the mean, and we use the constant of 10% decrease in standard deviation (STD) to reject the outliers. The initial 8 points in cluster $K$ are sorted in descending order by the distance from the mean $\mu$. When removing the first two points, the metric $m > 0.1$, but not when we remove the third, $m < 0.1$ thus cluster will be reduced to the most packed 6 points.

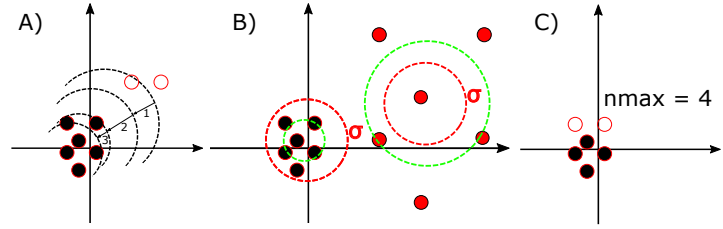

Fig. 10. A) Per cluster the outliers are removed according to predefined 10% decrease in STD. Only the first two points will be moved to the outliers because the 3rd point is close together with the other points; B) Global parameter of $\sigma_{thresh}$ in red and, per cluster, the estimated $\sigma$ in green, is either below or outside the expected range; C) Maximum number of points per cluster $n_{max}$, and $n_{pk} < n_{max}$ is the number of points in cluster

When the outliers are removed we want to reject the clusters that are overly spread for instance due to the drift. In our approach, the spread of particular cluster has to be below $\sigma_K < \sigma_{thresh}$ and it is computed according to Algorithm 3.

---

**Algorithm 3** Compute spread
---
1: $\mathbf{X} \leftarrow \mathbf{X} - \mu(\mathbf{X})$
2: $\mathbf{Y} \leftarrow \mathbf{Y} - \mu(\mathbf{Y})$          ▷ $cluster \leftarrow centertoZero(cluster)$
3: $\sigma_K = \sigma(K(:))$          ▷ $Eq.\ 10\ operating\ on\ cluster K$

---

The idea here is that the user selects a single cluster that with a big confidence corresponds to an isolated particle and passes the corresponding recommended limit of $\sigma_{thresh}$. The example of the thresholding by spread 10 B) shows that such a limit will be met by the set of black points but not by the red set. Finally, based on the geometric considerations outlined in Section 2 of the paper, we add a prior on the resultant amount of points that contribute to a single cluster. For a target particle diameter of the $d$, the amount of the zero-crossing points $(X, Y)_n \leq \frac{d}{\Delta y}$. Alternatively, one can perform numerical simulations where the line-by-line scanning of the focused spot is done through the range of particle sizes and by combining it with the estimates for the characteristic noise present in the technique assess the maximum amount of lines. Further research on this issue would be of interest; however, it is beyond the scope of this study.

Next, the two popular algorithms of K-means and DBSCAN are described and compared to become the cluster initializers.

**Classical K-means with prior**

Hyperparameters: $n_{max}, \sigma_{thresh}$

We modify the K-means [25, 45] such that it can accurately establish the isolated particles. For validation purposes, the density of spheres is crucial, thus the algorithm needs to overcome its inherent tendency to overestimate clusters. One difference to the original K-means is that we introduce the outliers. The outliers include: A) clusters with single particle; B) empty clusters; C) distant points previously included in a cluster. The option C) is treated by Algorithm 2. The second difference is conditioning of the assigned clusters. Clusters are considered to be valid if $n_{pk} < n_{max}$ (Figure 3 C)) and $\sigma'([X, Y]) < \sigma_{thresh}$, where $\sigma'([X, Y])$ is *spread* computed for the set of points, with a mean moved to the zero and the distance normalized to unity (Algorithm 3). After K-means convergence (one epoch), if there are points that fail on both conditions $n_{max}$ and $\sigma'([X, Y])$, they are passed through the K-means again. The algorithm stops if all points are assigned to either cluster or an outlier. In every epoch of the K-means, the optimal amount of clusters is defined from the elbow method based on the average of 3 random initializations. Hence, in our implementation, the K-means is described via following algorithm in pseudocode.

---

**Algorithm 4** K-means with prior
---
1: **procedure** KMEANSP($pasp, stdLimit, thresh$) ▷
   $pasp$ : $input\ points\ (X, Y), stdLimit : \sigma_{thresh}, thresh : n_{max}$
2:    **while** $size(pasp) > 0$ **do**
3:       $K \leftarrow optimalK$                                 ▷ Defined by elbow method
4:       $kmeans(pasp, K)$                              ▷ Apply classical K-means
5:       **for** $1 : K$ **do**                                         ▷ For each cluster
6:          **if** $Nremove \neq 0$ **then**           ▷ Remove outliers in cluster, Algorithm 2
7:             $set\ sp = sortedPoints(1 : Nremove)\ false$    ▷ Defining outliers
8:             $NoOutliers \leftarrow reverseSort(sp)$
9:          **end if**
10:         **if** $spread(cluster) > stdLimit$ **then**          ▷ Find sparse clusters
11:             $highSpread \leftarrow (spread(NoOutliers) > stdlimit)$
12:         **end if**
13:         **if** $size(cluster) > thresh$ **then**            ▷ Find dense clusters
14:             $ManyPoints \leftarrow (size(NoOutliers) > thresh)$
15:         **end if**
16:         $goodPoints \leftarrow \sim NoOutliers(highSpread\ \text{AND}\ ManyPoints)$
17:         $badPoints \leftarrow NoOutliers(highSpread\ \text{AND}\ ManyPoints)$
18:         **if** $size(goodPoints \leq 1)$ **then**     ▷ Remove single/zero point clusters
19:             *Remove isolated*
20:         **end if**
21:         $pasp \leftarrow badPoints$         ▷ Send not suitable points to new iteration
22:       **end for**
23:    **end while**
24: **end procedure**

---

**DBSCAN**

Hyperparameters: $n_{min}, \epsilon$

The density-based clustering algorithm (DBSCAN) [26, 46] has a straightforward advantage in taking care of the obscure points, such that all points that are not reachable from any other point are outliers or noise points. The two hyperparameters are inclusion radius $\epsilon$ and minimum number of points in the cluster $n_{min}$. The input for the $n_{min}$ is straightforward, such that it can be any number between $1 < n_{min} < n_{max}$. For the $\epsilon$ recommendation, we use the following routine:

- Normalize the complete data set of *middles* to the unity, such that $norm\mathbf{X} = \frac{\mathbf{X}}{max(\mathbf{X})}$ and $norm\mathbf{Y} = \frac{\mathbf{Y}}{max(\mathbf{Y})}$.

- Select the set of points that with a high confidence forms a cluster, via visual inspection, $confX = \{norm\mathbf{X}\}$ and $confY = \{norm\mathbf{Y}\}$. Center this cluster to the zero $centertoZero(confX, confY)$ (first two lines of Algorithm 3)

- Determine the average distance between points. Includes computation of Euclidian distance between each pair of observations in separately $\mathbf{X}$ and $\mathbf{Y}$ and taking average of each vector.

As a result of K-means or DBSCAN, one can pick up the converged clusters and either: A) Pull the features of *signalSize* specifically for the highest $V_{pk-pk}$ from corresponding pulses in a cluster; B) Average of the corresponding time-spans $\bar{\tau}$ of points in clusters (Eq. 12); C) The full amplitude of a signal itself $V_{pk-pk}$.

$$\bar{\tau} = \frac{\sum_{i=1}^{n_c} \tau_i}{n_c} \qquad (12)$$

Where $n_c$ is number of points assigned to the cluster. Correspondingly, the centroids of clusters are stored for the mapping of the particle positions.

**On computational complexity of algorithms. Comparing two algorithms**

The classical K-means algorithm has a complexity $O(TKn)$, where $n$ is the number of input points, $K$ is the desired number of clusters, and $T$ is the number of iterations needed for convergence. It is also observed that approximately $T \propto n$ [47]. Hence, the effective time complexity becomes $O(n^2)$. The K-means is a greedy algorithm since it can produce both empty and over-populated clusters. Another drawback is the large dependence on the initialization of cluster centers. As according to the quadratic time complexity, it should not be used in extremely large data applications [48]. Implementation of the K-means with prior in this paper has $O(n^2 \cdot log n)$ time complexity.

In the DBSCAN implementation, for each of the points of the input data, we have at most one region query. Thus, the average run time complexity of DBSCAN is query of $\log n$ times the amount of points $n$, $O(n \cdot log n)$.

**Sensitivity of K-means and DBSCAN algorithms**

Sensitivity analysis was used to explore how the accuracy of algorithms would change with slight variations in the hyperparameters. The green point in every plot represents the most preferred initial value that yields the highest accuracy, while the offset from this point defines the sensitivity (see Fig. 11).

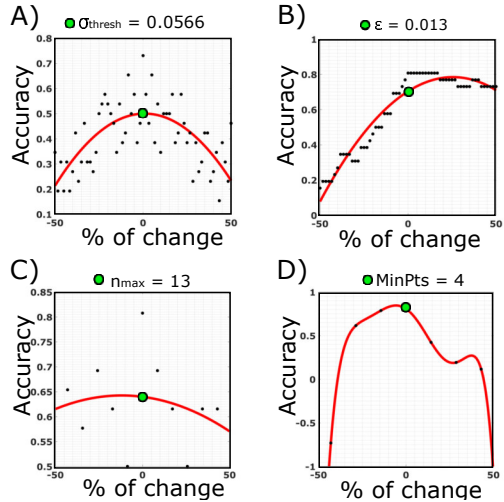

Fig. 11. Sensitivity analysis of isolated changes of hyperparameters for K-means and DBSCAN algorithms. The accuracy changes as a function of $\sigma_{thresh}$ and $\epsilon$, given fixed $n_{max} = 13$ and $MinPts = 4$ in A) and B) correspondingly and as a function of $n_{max}$ and $MinPts$, given $\sigma_{thresh} = 0.056$ and $\epsilon = 0.013$ in C) and D).

*Preparation of the samples*

Samples were prepared in a clean room class ISO 6 and we used high quality 1 inch wafers from Ultrasil. The general procedure for the sample preparation is outlined below:
- Clean UV/Ozone apparatus with IPA wipe and switch on for 15 minutes
- Prepare solution
- Place solution in ultrasonic bath
- Clean 1-inch Si wafer in UV Ozone for 5 minute
- Spin 0.5 ml solution on wafer @ 6100 RPM
- Place wafer in box

Solution for sample #1, 50 nm PSL: 3 droplets (Thermo scientific, Nanospheres, 3050A) in 0.5 *ml* demi water (from Mecrk Simplicity UV water purification system, applied in each recipe) 50 *μl* in 5 *ml* IPA (Sigma-Aldrich, 2-Propanol, anhydrous, catalogusnummer 278475-1L, applied in each recipe) under vigorous shaking.

Solution for sample #2, 100 nm PSL: 3 droplets (Thermo scientific, Nanosspheres, 3100A) in 0.5 *ml* demi water. Dilute 80 *μl* in 5 *ml* IPA.

Solution for sample #3, 60 and 80 nm PSL: 1 droplet 80 nm PSL dispersion (Thermo scientific, Nanospheres, 3080A) and 1 droplet 60 nm PSL dispersion (Thermo scientific, Nanospheres, 3060A) in 0.5 *ml* demi water. Dilute 70 *μl* in 5 *ml* IPA under vigorous shaking.

*The re-assignment of signal centroids by the K-means algorithm.*

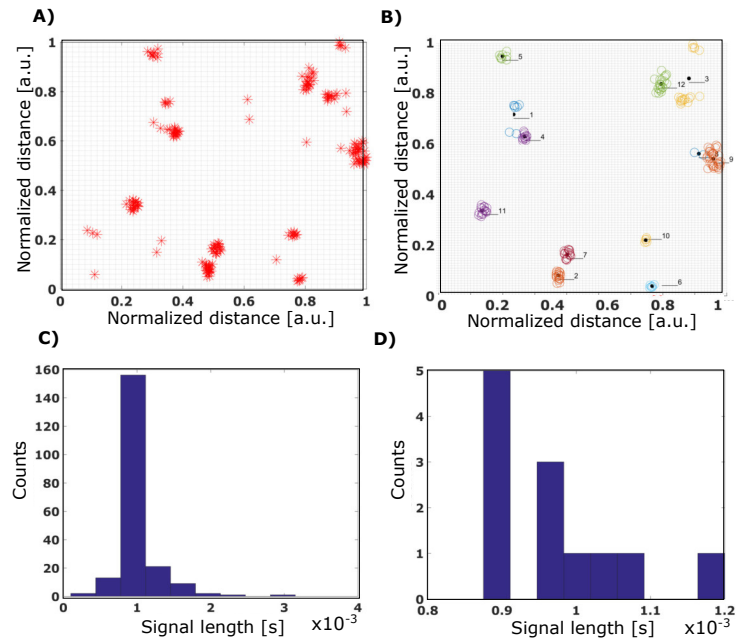

Fig. 12. Adapted K-means algorithm. A) The zero-crossings of the differential signals by the search algorithm and B) the corresponding isolated particles (after clustering) obtained by converged DBSCAN. The data set includes minor drift where one scanning line of $\Delta x = 25 \mu m$ takes 100 ms. Histograms obtained when C) all the particle-looking signals contribute to the histogram, and D) when the signals are clustered and only one centroid (with maximum $V_{pk-pk}$) is assigned to represent the particle.

In addition to the better performing algorithm of DBSCAN presented in Section 5.3, we demonstrate the output of the modified K-means algorithm 12 B). The difference with the DBSCAN algorithm is a tendency to merge the clusters that would easily be separated by the human eye. Such cluster can be seen as cluster n. 3, which contains two separable groups of points. The solution for this problem can be to re-initialize the algorithm multiple times until the cluster is assigned correctly. Nevertheless, it is more informative to present average initialization result. The algorithm is capable of separating three classes of particles, as shown in 12 D), which is much better than the result of using all the detected signals 12 C).

## Funding

Nederlandse Organisatie voor Wetenschappelijk Onderzoek; 501100003246; 14660, High Tech Systems and Materials Research Program, 14660, Applied and Technical Sciences division (TTW).

## Acknowledgments

We gratefully acknowledge the help provided by TNO in the fabrication of samples. Dmytro Kolenov acknowledges the High Tech Systems and Materials Research Program with Project n. 14660, financed by the Netherlands Organisation for Scientific Research (NWO), Applied and Technical Sciences division (TTW) for funding this research.

## Disclosures

The authors declare no conflicts of interest.